\documentclass[amssymb, amsmath, reprint, pra, showpacs]{revtex4-1}

\usepackage{microtype}
\usepackage{graphicx}
\usepackage{xcolor}
\usepackage{siunitx}
\usepackage{hyperref}
\usepackage{cleveref}
\usepackage{braket}
\usepackage{color}
\usepackage{dsfont}
\usepackage{pbox}
\usepackage{relsize}
\usepackage{diagbox}
\usepackage{enumerate}

\usepackage{tikz}
\usetikzlibrary{decorations.pathreplacing}

\definecolor{niceblue}{HTML}{337ab7}
\definecolor{nicegreen}{HTML}{5cb85c}

\DeclareMathOperator{\tr}{tr}
\newcommand{\dd}{\text{d}}

\newcommand{\ketbra}[2]{|#1\rangle\langle#2|}

\begin{document}

\title{Necessary and sufficient conditions for macroscopic realism from\\ quantum mechanics}
\date{\today}

\author{Lucas Clemente}
\author{Johannes Kofler}
\affiliation{Max-Planck-Institute of Quantum Optics, Hans-Kopfermann Str.\ 1, 85748 Garching, Germany}
\email{lucas.clemente@mpq.mpg.de, johannes.kofler@mpq.mpg.de}


\begin{abstract}
Macroscopic realism, the classical world view that macroscopic objects exist independently of and are not influenced by measurements, is usually tested using Leggett-Garg inequalities. Recently, another necessary condition called no-signaling in time (NSIT) has been proposed as a witness for non-classical behavior. In this paper, we show that a combination of NSIT conditions is not only necessary but also sufficient for a macrorealistic description of a physical system. Any violation of macroscopic realism must therefore be witnessed by a suitable NSIT condition. Subsequently, we derive an operational formulation for NSIT in terms of positive operator-valued measurements and the system Hamiltonian. We argue that this leads to a suitable definition of ``classical'' measurements and Hamiltonians, and apply our formalism to some generic coarse-grained quantum measurements.
\end{abstract}

\maketitle

\section{Introduction}
\label{sec:introduction}

Whether or not the laws of quantum mechanics are universally valid and hold on the level of macroscopic objects, is still an open question in the physics community. Some believe that the issue will be settled in favor of quantum theory by the experimental demonstration of Schr\"odinger cat-like states \cite{Schrodinger:1935kq}. Others hold that some physical mechanism, altering the laws of quantum mechanics \cite{Ghirardi:1986ep, Diosi:1989cv, Penrose:1998ek}, guarantees a fully classical world on the macroscopic level.

In 1985, Leggett and Garg \cite{Leggett:1985bl} have put forward macroscopic realism (or macrorealism), a world view encompassing all physical theories which enforce that macroscopic properties of macroscopic objects exist independently of and are not influenced by measurement. While setups such as superconducting devices, heavy molecules, and quantum-optical systems are promising candidates in the race towards an experimental violation of macrorealism, non-classical effects have so far only been observed for microscopic objects or microscopic properties of larger objects \cite{Friedman:2000gs, Julsgaard:2001by, PalaciosLaloy:2010ih, Gerlich:2011go, Dressel:2011hh, Goggin:2011iw, Fedrizzi:2011ji, Waldherr:2011km, Knee:2012cg, Lvovsky:2013gh, Sekatski:2014kc, Ghobadi:2014eq, Asadian:2014fw, Robens:2015ba}. However, a genuine violation of macroscopic realism---with its reference to macroscopically distinct states---requires using solely measurements of macroscopically coarse-grained observables. Note that there are several approaches to quantifying the ``macroscopicity'' of quantum states and measurements \cite{Dur:2002fr, Korsbakken:2007bc, Lee:2011gi, Nimmrichter:2013gc, Mraz:2014jg, Jeong:2015gp, Laghaout:2015kt, Frowis:2015iq}. It is also known that usually the restriction to such coarse-grained (``classical'') measurements alone already leads to the emergence of classicality \cite{Kofler:2007bd}, unless a certain type of (``non-classical'') Hamiltonian is governing the object's time evolution \cite{Kofler:2008dz}. Recent investigations have confirmed the intuition that these Hamiltonians are hard to engineer and require a very high control precision in the experimental setup \cite{Wang:2013ir, Jeong:2014hw, Sekatski:2014ej}.

A quantum violation of macrorealism (MR) is usually witnessed by the violation of a Leggett-Garg inequality (LGI), which is composed of temporal correlations between sequential measurements of an object undergoing time evolution. Recently, following earlier works \cite{Kofler:2008dz, Clifton:1991ta, Foster:1991jz, Benatti:1994fy}, another necessary condition for MR called no-signaling in time (NSIT) was proposed \cite{Kofler:2013hb}. It can be regarded as a statistical version of the non-invasive measurability postulate.

In \cref{sec:nim}, we start with the discussion of various instances of NSIT and show that in the correct combination they form a sufficient condition for a macrorealistic description (at a given set of possible measurement times). We also demonstrate that it is impossible to establish such a sufficient condition for a macrorealistic description by combining LGIs involving two-time measurements. Subsequently, in \cref{sec:nsit}, we derive an operational condition for NSIT, based on (projective and non-projective) measurement operators and the system Hamiltonian. In \cref{sec:classicality}, we use these results to define the classicality of measurements based on a reference set of a-priori classical operators and to characterize the classicality of Hamiltonians. Finally, in \cref{sec:examples}, we apply our formalism to measurements of coherent states, quadratures, and Fock states, and quantify their invasiveness as a function of their coarse-graining.

\section{Non-invasive measurements}
\label{sec:nim}

Let us start with the definition of macrorealism, consisting of the following postulates \cite{Leggett:2002dk}: ``(1) \emph{Macrorealism per se}. A macroscopic object which has available to it two or more macroscopically distinct states is at any given time in a definite one of those states. (2) \emph{Non-invasive measurability}. It is possible in principle to determine which of these states the system is in without any effect on the state itself or on the subsequent system dynamics. (3) \emph{Induction}. The properties of ensembles are determined exclusively by initial conditions (and in particular not by final conditions).''

In the following, we will first show that a strong reading of non-invasive measurability implies macrorealism per se (\cref{sub:mrps}). Then we will present various necessary conditions (\cref{sub:necessary-conditions}) and a set of sufficient conditions (\cref{sub:sufficient-conditions}) for a macrorealistic description.

\subsection{Macrorealism per se following from strong non-invasive measurability}
\label{sub:mrps}

In this subsection, we assume that the state space of a macroscopic object is split into macroscopically distinct \emph{non-overlapping} states (macrostates). Consider a macro-observable $Q(t)$ with a one-to-one mapping between its values and the macrostates. Further consider measurements of the macro-observable that enforce a definite post-measurement macrostate and report the corresponding value as the outcome.

Macrorealism per se (MRps) is fulfilled if $Q(t)$ has a definite value at all times $t$, prior to and independent of measurement:
\begin{equation}\label{eq:mrps}
  \forall t\!: \exists ~\text{definite}~ Q(t).
\end{equation}
Probabilistic predictions for $Q(t)$ are merely due to ignorance of the observer. Even in cases where $Q(t)$ evolves unpredictably (e.g.\ in classical chaos) or even indeterministically, it is still assumed to have a definite value at all times.

On top of MRps, the assumption of non-invasive measurability (NIM) in principle allows a measurement at every instant of time, revealing the macrostate without disturbance. NIM guarantees that
\begin{equation}\label{eq:nim-hidden}
  \forall t\!: Q(t) = Q_H(t),
\end{equation}
where $H$ denotes the history of past non-invasive measurements on the system: In order for measurements to be non-invasive, the time evolution of $Q$ must not depend on the history of the experiment \footnote{Let us now assume the existence of hidden parameters $\lambda(t)$ that define all physical properties. MRps is fulfilled if the macro-observable is a deterministic function $Q = Q(\lambda(t))$. There are two conceivable scenarios: (i) \emph{Deterministic time evolution} of $\lambda$, causing deterministic time evolution of the macro-observable $Q(\lambda(t))$. (ii) \emph{Stochastic time evolution} of $\lambda$, where some intrinsic randomness generates random jumps in $\lambda$. We still have a deterministic dependency $Q(\lambda)$, but $Q(\lambda(t))$ \emph{appears} stochastic. In both cases MRps is fulfilled, since the system is in a single macrostate, as described by $Q = Q(\lambda(t))$, at all times. The condition for NIM then reads $Q(\lambda(t)) = Q(\lambda_H(t))$, where $\lambda_H(t)$ are the hidden parameters after a history $H$ of non-invasive measurements.}. Note that all non-invasive measurements are repeatable, i.e.\ when performing the same measurement immediately again, the same outcome is obtained with probability 1.

In the literature, NIM is often treated as a necessary condition for macrorealism per se. It is argued that NIM is ``so natural a corollary of [MRps] that the latter is virtually meaningless in its absence'' \cite{Leggett:2002dk}. As some others before \cite{Kofler:2013hb, Bacciagaluppi:2014ue, Maroney:2014ws-arxiv}, we do not adhere to this position. A counter example to the statement $\text{MRps} \Rightarrow \text{NIM}$ is given by the de Broglie--Bohm theory, where measurements are invasive, as they affect the guiding field and thus the subsequent (position) state, but MRps is fulfilled, as the (position) state is well-defined at all times. In fact, we now argue that there exist two different ways of reading the postulate of NIM in \cite{Leggett:2002dk}:
\begin{itemize}
  \item \emph{Weak NIM}. Given a macroscopic object is in a definite one of its macrostates, it is possible to determine this state without any effect on the state itself or on the subsequent system dynamics.
  \item \emph{Strong NIM (sNIM)}. It is always possible to measure the macrostate of an object without any effect on the state itself or on the subsequent system dynamics.
\end{itemize}
Let us now argue that sNIM actually implies MRps. Assuming sNIM, a hypothetical non-invasive measurement can be performed at every instant of time, determining the value of the macro-observable $Q$. Due to its non-invasive nature, $Q$ must have had a definite value already before the measurement. This ensures that $Q$ has a definite value at all times, giving rise to a ``trajectory'' $Q(t)$. Therefore,
\begin{equation}\label{eq:nim-mrps}
  \text{sNIM} \Rightarrow \text{MRps}.
\end{equation}
Another way of establishing this implication is the following: Assume that MRps fails, i.e.\ the object is not in a definite macrostate. A measurement leaves the object in a definite macrostate, creating a definite state out of an indefinite one, and therefore does not satisfy sNIM. We thus have $\lnot \text{MRps} \Rightarrow \lnot \text{sNIM}$, which is equivalent to \eqref{eq:nim-mrps}.

Note that \eqref{eq:nim-mrps} holds even if sNIM is made less stringent, allowing measurements to change the subsequent time evolution, while still determining the macrostate.

In this paper, we implicitly assume induction (the arrow of time) \cite{Leggett:2002dk} and freedom of choice concerning the initial states and measurement times (including whether a measurement takes place at all). Then, sNIM alone is sufficient for macrorealism, and by extension, for testable conditions such as the Leggett-Garg inequalities or no-signaling in time \cite{Kofler:2013hb}:
\begin{equation}
  \text{sNIM} \Leftrightarrow \text{MRps} \land \text{NIM} \Leftrightarrow \text{MR} \Rightarrow \text{LGI, NSIT}.
\end{equation}

Let us remark that NIM is in general not as strongly physically motivated as the assumption of locality in Bell's theorem. The so-called ``clumsiness loophole'' allows violations of NIM to be attributed to imperfections of the measurement apparatus instead of genuine quantum effects. This loophole can be addressed using ideal negative measurements \cite{Leggett:1985bl} or more involved protocols \cite{Wilde:2011ip}.

\subsection{Necessary conditions for macrorealism}
\label{sub:necessary-conditions}

\begin{figure}[t]
  \begin{tikzpicture}
    \draw [thick, ->] (-3, 0) -- (3, 0);

    \node[right] at (3, 0) {$t$};
    \node at (-2, -0.4) {$t_0$};
    \node at (0, -0.4) {$t_1$};
    \node at (2, -0.4) {$t_2$};
    \draw (-2, 0.1) -- (-2, -0.1);
    \draw (0, 0.1) -- (0, -0.1);
    \draw (2, 0.1) -- (2, -0.1);
    \node[anchor=base] at (-1, 0.3) {$\hat U_{01}$};
    \node[anchor=base] at (1, 0.3) {$\hat U_{12}$};
    \node[anchor=base] at (-2.2, 0.3) {$\hat \rho_0$};

    \newcommand{\tCondOverTime}[5]{
      \node[left] at (-3, -1-0.5*#1) {#5};
      \draw [->] (-3, -1-0.5*#1) -- (3, -1-0.5*#1);
      \ifx&#2&\else\draw [fill=#2] (-2, -1-0.5*#1) circle [radius=0.07];\fi
      \ifx&#3&\else\draw [fill=#3] (0, -1-0.5*#1) circle [radius=0.07];\fi
      \ifx&#4&\else\draw [fill=#4] (2, -1-0.5*#1) circle [radius=0.07];\fi
    }

    \tCondOverTime{0}{lightgray}{lightgray}{lightgray}{LGI$_{012}$}
    \tCondOverTime{1}{white}{black}{}{NSIT$_{(0)1}$}
    \tCondOverTime{2}{}{white}{black}{NSIT$_{(1)2}$}
    \tCondOverTime{3}{white}{}{black}{NSIT$_{(0)2}$}
    \tCondOverTime{4}{black}{white}{black}{NSIT$_{0(1)2}$}
    \tCondOverTime{5}{white}{black}{black}{NSIT$_{(0)12}$}
    \tCondOverTime{6}{black}{white}{black}{NIC$_{0(1)2}$}
  \end{tikzpicture}

  \caption{\label{fig:lgi-nsit}Different necessary conditions for MR in a system with possible measurements at three points in time. Black filled circles denote measurements that always take place, white filled circles measurements that may or may not be performed. A pair of measurements is always performed for the LGI, shown with gray filled circles.}
\end{figure}
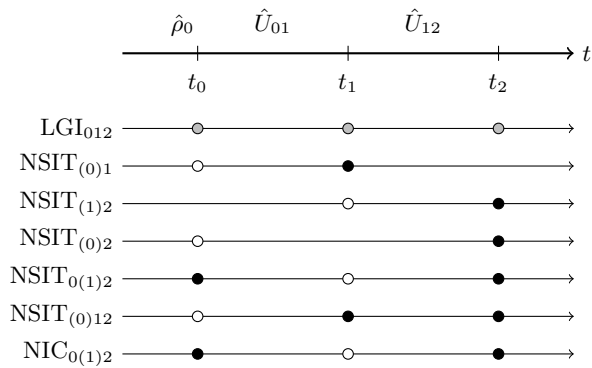

The relationship between LGI and NSIT has previously been discussed in the literature for a number of example systems \cite{Kofler:2008dz, Saha:2014un-arxiv, Kofler:2013hb, Maroney:2014ws-arxiv}. Here we consider the archetypal setup depicted in \cref{fig:lgi-nsit}: A system starting in the initial state $\hat \rho_0$ evolves under unitary $\hat U_{01}$ from $t_0$ to $t_1$, and under unitary $\hat U_{12}$ from $t_1$ to $t_2$. During the evolution, dichotomic measurements may be performed at times $t_i$ for $i \in \lbrace 0, 1, 2 \rbrace$. Let us call the outcomes of these measurements $Q_i \in \lbrace -1, +1 \rbrace$, and define the correlations $C_{ij} = \langle Q_i Q_j \rangle$. Then, the simplest LGI reads
\begin{equation}\label{eq:lgi}
  \text{LGI}_{012}\!: C_{01} + C_{12} - C_{02} \leq 1.
\end{equation}
There exist many other Leggett-Garg inequalities involving more than three possible measurement times or more than two outcomes (for a recent review see \cite{Emary:2014ck}). Quantum mechanical experiments are able to violate ineq.\ \eqref{eq:lgi} up to $1.5$ for a qubit and, as shown in \cite{Budroni:2014fc}, up to the algebraic maximum $3$ for higher-dimensional systems still using dichotomic measurements $Q_i = \pm 1$.

On the other hand, NSIT$_{(i)j}$ is a statistical version of \cref{eq:nim-hidden}, requiring that the outcome probabilities $P_j(Q_j)$ of result $Q_j$ measured at time $t_j$ are the same, no matter whether or not a measurement was performed at some earlier time $t_i < t_j$:
\begin{equation}
  \label{eq:NSIT}\text{NSIT}_{(i)j}\!: P_j(Q_j) = P_{ij}(Q_j) \equiv \sum_{Q_i'} P_{ij}(Q_i', Q_j).
\end{equation}
Note that the probability distributions on both sides of the equation, $P_{i}$ and $P_{ij}$, correspond to \emph{different} physical experiments: While $P_j$ is established by measuring only at $t_j$, $P_{ij}$ is obtained by measuring both at $t_i$ and $t_j$. Unlike in the LGI in \eqref{eq:lgi}, one is not limited to only two outcomes. If it is the later measurement at $t_j$ which may or may not be performed, NSIT$_{i(j)}$ is an instance of the arrow of time and is therefore fulfilled by both macrorealism and quantum mechanics.

While NSIT$_{(1)2}$ is a promising condition that is usually able to detect violations of MR more reliably than LGI$_{012}$ \cite{Kofler:2013hb, Saha:2014un-arxiv}, it fails for particular initial states, where the invasiveness is able to ``hide'' in the statistics of the experiment (see the discussion below). We can however make NSIT$_{(1)2}$ robust against such cases, by always performing a measurement at $t_0$. We call the resulting condition
\begin{equation}
  \begin{split}
    \text{NSIT}_{0(1)2}\!: P_{02}(Q_0, Q_2) &= P_{012}(Q_0, Q_2) \\
                                                    &\equiv \sum_{Q_1'} P_{012}(Q_0, Q_1', Q_2).
  \end{split}
\end{equation}
NSIT$_{0(1)2}$ alone is not sufficient for LGI$_{012}$. Hence, we also introduce the condition
\begin{equation}
  \begin{split}
    \text{NSIT}_{(0)12}\!: P_{12}(Q_1, Q_2) &= P_{012}(Q_1, Q_2) \\
                                                  &\equiv \sum_{Q_0'} P_{012}(Q_0', Q_1, Q_2).
  \end{split}
\end{equation}
As was recently shown in \cite{Maroney:2014ws-arxiv}, a combination of NSIT$_{(0)12}$, NSIT$_{0(1)2}$ and the arrow of time (AoT) is sufficient for LGI$_{012}$:
\begin{equation}\label{eq:nsit-lgi}
  \text{NSIT}_{0(1)2} \land \text{NSIT}_{(0)12} \land \text{AoT} \Rightarrow \text{LGI}_{012}.
\end{equation}
The inverse is not true, and moreover the left-hand side is not sufficient for macrorealism (see discussion below).

We further remark that one can also write a condition similar to NSIT$_{0(1)2}$ in a more intuitive form that we call non-invaded correlations (NIC),
\begin{equation}
  \text{NIC}_{0(1)2}\!: C_{02} = C_{02|1},
\end{equation}
where $C_{02|1}$ denotes the correlation $\langle Q_0 Q_2 \rangle$ given that an additional measurement was performed at $t_1$. It is shown in \cref{appendix:nsit-nic} that NIC$_{0(1)2}$ follows from NSIT$_{0(1)2}$.

Fig.\ \ref{fig:lgi-nsit} presents a graphical summary of the conditions that have been discussed in this section.

\subsection{NSITs as sufficient conditions for macrorealism}
\label{sub:sufficient-conditions}

In the following, we will show that the combination of various NSIT conditions and the arrow of time (AoT) guarantees the existence of a unique global probability distribution $P_{012}(Q_0, Q_1, Q_2)$, which is equivalent to macrorealism evaluated at $t_0, t_1, t_2$. Let us start by writing all single-measurement probabilities in terms of $P_{012}$. Once again, note that joint probabilities $P$ with different subscripts correspond to different experimental setups (e.g.\ $P_2(Q_2)$ is obtained by measuring only at $t_2$, while $P_{12}(Q_1, Q_2)$ is obtained by measuring at times $t_1$ and $t_2$):
\begin{equation}\label{eq:mr-p2}
  P_2(Q_2) = \sum_{Q_1'} P_{12}(Q_1', Q_2) = \sum_{Q_0'} \sum_{Q_1'} P_{012}(Q_0', Q_1', Q_2),
\end{equation}
where we have used NSIT$_{(1)2}$ for the first equality and NSIT$_{(0)12}$ for the second one. Furthermore,
\begin{equation}\label{eq:mr-p1}
  P_1(Q_1) = \sum_{Q_2'} P_{12}(Q_1, Q_2') = \sum_{Q_0'} \sum_{Q_2'} P_{012}(Q_0', Q_1, Q_2'),
\end{equation}
where for the first equality we assumed AoT [i.e.\ $Q_i$ are (statistically) independent of $Q_j$ for $j>i$], and NSIT$_{(0)12}$ for the second one. Moreover, we see that
\begin{equation}
  P_0(Q_0) = \sum_{Q_1'} \sum_{Q_2'} P_{012}(Q_0, Q_1', Q_2'),
\end{equation}
where AoT was used twice. Next, the pairwise joint probability functions can be constructed:
\begin{equation}
  P_{01}(Q_0, Q_1) = \sum_{Q_2'} P_{012}(Q_0, Q_1, Q_2')
\end{equation}
follows from AoT\@. Using NSIT$_{0(1)2}$ one obtains
\begin{equation}
  P_{02}(Q_0, Q_2) = \sum_{Q_1'} P_{012}(Q_0, Q_1', Q_2).
\end{equation}
Finally, using NSIT$_{(0)12}$, we obtain
\begin{equation}
  P_{12}(Q_1, Q_2) = \sum_{Q_0'} P_{012}(Q_0', Q_1, Q_2).
\end{equation}

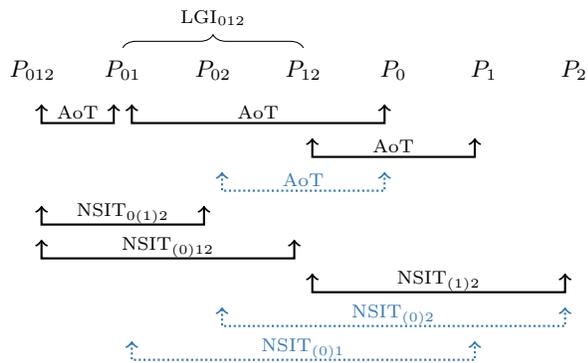
\begin{figure}[t]
  \begin{tikzpicture}[xscale=1.2]
    \def\dy{-0.45} 
    \def\ady{-0.25} 
    \def\tdy{-0.1} 
    \def\adx{0.1} 

    \draw [decorate,decoration={brace,amplitude=05pt}] (1, 0.3) -- (3, 0.3);
    \node at (2, 0.7) {\scriptsize LGI$_{012}$};

    \node at (0, 0) {$P_{012}$};
    \node at (1, 0) {$P_{01}$};
    \node at (2, 0) {$P_{02}$};
    \node at (3, 0) {$P_{12}$};
    \node at (4, 0) {$P_0$};
    \node at (5, 0) {$P_1$};
    \node at (6, 0) {$P_2$};

    \newcommand{\condArrow}[5][black]{%
      \draw[thick,<->,color=#1] (#2+0.1, #4*\dy) -- (#2+0.1, #4*\dy+\ady) -- (#3-0.1, #4*\dy+\ady) -- (#3-0.1, #4*\dy);
      \node[color=#1] at (#2/2+#3/2, #4*\dy+\tdy) {\scriptsize #5};
    }

    \condArrow{0}{1}{1}{AoT}
    \condArrow{1}{4}{1}{AoT}
    \condArrow{3}{5}{2}{AoT}
    \condArrow[niceblue,densely dotted]{2}{4}{3}{AoT}

    \condArrow{0}{2}{4}{NSIT$_{0(1)2}$}
    \condArrow{0}{3}{5}{NSIT$_{(0)12}$}

    \condArrow{3}{6}{6}{NSIT$_{(1)2}$}
    \condArrow[niceblue,densely dotted]{2}{6}{7}{NSIT$_{(0)2}$}

    \condArrow[niceblue,densely dotted]{1}{5}{8}{NSIT$_{(0)1}$}

  \end{tikzpicture}
  \caption{\label{fig:mr}(Color online) Different combinations of NSIT and AoT conditions are sufficient for guaranteeing that all probability distributions $P_i, P_{ij}$ are the marginals of a unique global probability distribution $P_{012}$. There are multiple ways of obtaining a sufficient set. The black arrows correspond to one particular choice, and additional conditions are printed for completeness in blue. Note that the existence of a classical explanation for the pairwise joint probabilities $P_{ij}$ is sufficient for fulfilling LGI$_{012}$, but not for MR$_{012}$.}
\end{figure}

We have thus shown that there exists a combination of NSIT conditions, whose fulfillment guarantees that all probability distributions in any experiment can be written as the marginals of a unique global probability distribution $P_{012}(Q_0, Q_1, Q_2)$. This is equivalent to the existence of a macrorealistic model for measurements at times $t_0, t_1, t_2$ (MR$_{012}$). Note that while MR$_{012}$ cannot prove the world view of MR in general, it implies that no experimental procedure (with measurements at $t_0, t_1, t_2$) can detect a violation of MR\@. Let us now write a \emph{necessary and sufficient} condition for MR$_{012}$,
\begin{equation}\label{eq:mr012}
  \text{NSIT}_{(1)2} \land \text{NSIT}_{0(1)2} \land \text{NSIT}_{(0)12} \land \text{AoT} \Leftrightarrow \text{MR}_{012}.
\end{equation}
This set of conditions is not unique: We can e.g.\ substitute NSIT$_{(1)2}$ by NSIT$_{(0)2}$, as can easily be seen from a graphical representation of all conditions in \cref{fig:mr}. We remark that even the combination of all two-time NSIT conditions, $\text{NSIT}_{(0)1} \land \text{NSIT}_{(1)2} \land \text{NSIT}_{(0)2}$, is sufficient neither for MR$_{012}$ nor for LGI$_{012}$. Note that LGIs only test for non-classicalities of the pairwise joint probability distributions. A smaller set of conditions is therefore sufficient for fulfilling all LGIs using two-time correlation functions or probabilities [such as ineq.\ \eqref{eq:lgi} or the so-called Wigner LGIs \cite{Saha:2014un-arxiv}], see expression \eqref{eq:nsit-lgi}.

To illustrate these conditions for a qubit, in \cref{table:qubit} we show the individual conditions evaluated for a Mach-Zehnder setup (reflectivities $R_1, R_2$, phase plate $\varphi$ in one arm) with arbitrary initial state and time evolution. The three possible measurements are which-path measurements before the first beam splitter ($t_0$), between the two beamsplitters ($t_1$), and after the second beamsplitter ($t_2$), respectively. We can easily find cases where LGI$_{012}$ is always fulfilled, but various NSIT conditions still witness a violation of MR, e.g.\ for $R_1 = R_2 = 1/2, \varphi \neq (n+1/2)\pi$. As discussed above, it is possible for LGI$_{012}$ to be violated with NSIT$_{(1)2}$ fulfilled, e.g.\ for $R_1=1/4, R_2=3/4, q=1/2, \varphi=\pi$. For mixed initial states, NSIT$_{0(1)2}$ reduces to the condition $\varphi = (n+1/2) \pi$ with $n \in \mathbb N_0$ and is sufficient for MR$_{012}$, as no interference is possible in this case. For general superposition states, NSIT$_{(0)12}$ can be violated with NSIT$_{0(1)2}$ fulfilled. Moreover, NSIT conditions still allow detecting violations of MR if $R_1 = 0,1$ or $R_2 = 0,1$.

\begin{table*}[t]
  \newcommand{\fulfilled}{\checkmark}
  \renewcommand*{\arraystretch}{1.4}
  \begin{ruledtabular}
    \begin{tabular}{ll|c|c|c|c}
      & & LGI$_{012}$ & NSIT$_{(1)2}$ & NSIT$_{0(1)2}$ & NSIT$_{(0)12}$ \\\hline
      $\hat \rho_\text{mix}\!:$ & $R_1 = R_2 = \frac{1}{2}$ & \fulfilled & $q = \frac{1}{2}$ or $\varphi = (n+\frac{1}{2}) \pi$ & $\varphi = (n+\frac{1}{2}) \pi$ & \fulfilled \\
                             & $R_1 = \frac{1}{4}, R_2 = \frac{3}{4}$ & $1 + 3 \cos \varphi \geq 0$ & $q = \frac{1}{2}$ or $\varphi = (n+\frac{1}{2}) \pi$ & $\varphi = (n+\frac{1}{2}) \pi$ & \fulfilled \\
                             & $R_1, R_2$                             & $R_1 + \alpha \cos \varphi - R_1 R_2 \geq 0$ & $q = \frac{1}{2}$ or $\varphi = (n+\frac{1}{2}) \pi$ or $\alpha = 0$ & $\varphi = (n+\frac{1}{2}) \pi$ or $\alpha = 0$ & \fulfilled \\ \hline
      $\hat \rho_\text{sup}\!:$ & $R_1 = R_2 = \frac{1}{2}$ & \fulfilled & $2 q \cos\varphi = \cos\varphi + 2 \operatorname{Re}(c) \sin \varphi$ & $\varphi = (n+\frac{1}{2}) \pi$ & $c \in \mathbb R$ \\
                                & $R_1 = \frac{1}{4}, R_2 = \frac{3}{4}$ & $1 + 3 \cos \varphi \geq 0$ & $[\ast]$ & $\varphi = (n+\frac{1}{2}) \pi$ & $c \in \mathbb R$ \\
                             & $R_1, R_2$                             & $R_1 + \alpha \cos \varphi - R_1 R_2 \geq 0$ & $[\ast\ast]$ & $\varphi = (n+\frac{1}{2}) \pi$ or $\alpha = 0$ & $c \in \mathbb R$ or $R_1 = 0,1$
    \end{tabular}
  \end{ruledtabular}
\caption[]{\label{table:qubit}Different necessary conditions for macrorealism evaluated for a Mach-Zehnder (qubit) experiment \cite{Reck:1994dz}. The reflectivity of the first beamsplitter is $R_1$, and of the second one is $R_2$. In one path of the interferometer, a phase $\varphi$ is added. Which-path measurements may be performed before, between and after the beamsplitters. The initial states are $\hat \rho_\text{mix} = \left(\begin{smallmatrix}q&0\\0&1-q\end{smallmatrix}\right)$ and $\hat \rho_\text{sup} = \left(\begin{smallmatrix}q&c\\c^\ast&1-q\end{smallmatrix}\right)$. The symbol ``\fulfilled'' means that the condition holds for all values of the free parameters. For brevity, $\alpha \equiv \sqrt{R_1 R_2 (1-R_1) (1-R_2)}$. Equation $[\ast]$ reads $(2 i \sqrt{3} c+6 q-3) \cos \varphi -2 i \sqrt{3} \operatorname{Re}(c) (\cos \varphi -2 i \sin \varphi )=0$, equation $[\ast\ast]$ reads $\cos \varphi [(2 q-1) \alpha +i c (1-2 R_1) \sqrt{-(R_2-1) R_2}]+i \sqrt{-(R_2-1) R_2} \operatorname{Re}(c) [(2 R_1-1) \cos \varphi +i \sin \varphi ]=0$. See main text for discussion.}
\end{table*}

\section{NSIT for quantum measurements}
\label{sec:nsit}

In the following, we will look at NSIT$_{(0)T}$ in an archetypal quantum experiment. A system has been prepared at $t = 0$ in an initial state $\hat \rho_0$. Then, at $t = 0$, a POVM $\lbrace \hat A_a^\dagger \hat A_a \rbrace_a$ with outcomes $a$ is carried out. After the measurement, the system evolves according to a unitary $\hat U = e^{-i \hat H t}$. At time $t = T$ a second, possibly different POVM $\lbrace\hat B_b^\dagger \hat B_b\rbrace_b$ with outcomes $b$ is performed.

\begin{figure}[tb]
  \begin{tikzpicture}
    \draw [thick, ->] (-3, 0.5) -- (3, 0.5);
    \draw [thick, ->] (-3, -0.5) -- (3, -0.5);

    \draw [fill] (-2, 0.5) circle [radius=0.07];
    \draw (-2, -0.6) -- (-2, -0.4);
    \node at (-2, 0) {$t = 0$};
    \node at (-2, 1) {$\hat A_a$};

    \draw [fill] (2, 0.5) circle [radius=0.07];
    \draw [fill] (2, -0.5) circle [radius=0.07];
    \node at (2, 0) {$t = T$};
    \node at (2, 1) {$\hat B_b$};
    \node at (2, -1) {$\hat B_b$};
    \node [left] at (-3.1, -0.5) {$P_{\hat B}(b)$};
    \node [left] at (-3.1, 0.5) {$P_{\hat B|\hat A}(b)$};
    \node [right] at (3.1, -0.5) {$t$};
    \node [right] at (3.1, 0.5) {$t$};

    \node at (0, 0) {$\hat H$};
  \end{tikzpicture}

  \caption{\label{fig:setup}A system evolves from $t = 0$ to $t = T$ under Hamiltonian $\hat H$. In the first setup measurements $\hat A_a^\dagger \hat A_a$ and $\hat B_b^\dagger \hat B_b$ are performed at $t = 0$ and $t = T$, respectively, and in the second setup only a final measurement $\hat B_b^\dagger \hat B_b$ is performed.}
\end{figure}
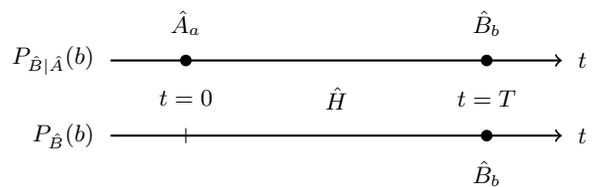

To determine the effect of the first measurement $\hat A_a^\dagger \hat A_a$ on the system's state and its subsequent dynamics, we will compare the results of the final measurement with a different experiment, where no measurement was performed at $t = 0$ (or, equivalently, a measurement $\hat A_a = \mathds 1$ was performed). The two setups are shown in \cref{fig:setup}.

The probabilities for obtaining outcome $b$ in the second and first setup are called $P_{\hat B}(b)$ and $P_{\hat B | \hat A}(b)$, respectively. They can be calculated as
\begin{align}
  \label{eq:P-unmeasured}
  P_{\hat B}(b) &= \tr ( \hat B_b \hat U_T \hat \rho_0 \hat U_T^\dagger \hat B_b^\dagger) \\
  \label{eq:P-measured}
  P_{\hat B | \hat A}(b) &= \int \dd a\, \tr ( \hat B_b \hat U_T \hat A_a \hat \rho_0 \hat A_a^\dagger \hat U_T^\dagger \hat B_b^\dagger),
\end{align}
with the integral replaced by a sum if the number of outcomes is countable. NSIT$_{(0)T}$ is fulfilled if the test measurement has no detectable effect on the system, i.e.\ if $P_{\hat B} = P_{\hat B | \hat A}$:
\begin{equation}
  \label{eq:NSIT-operators}
  \forall b\!: \tr ( \hat B_b \hat U_T \hat \rho_0 \hat U_T^\dagger \hat B_b^\dagger) = \int \dd a\, \tr ( \hat B_b \hat U_T \hat A_a \hat \rho_0 \hat A_a^\dagger \hat U_T^\dagger \hat B_b^\dagger).
\end{equation}
Note that the equality sign in \cref{eq:NSIT-operators} will often be fulfilled only approximately, even by non-invasive measurements. In practice, one can choose from a variety of error measures and corresponding reasonable error thresholds. However, to simplify notation, we will continue to use the equality sign in the following calculations.

\subsection{NSIT without time evolution}
\label{sub:nsit-no-time}

Let us start by considering the case $T = 0$ (NSIT$_{(0)0}$), i.e.\ the final measurement is performed immediately after the test measurement. In this setup, NSIT can be regarded as a case of joint measurability, a condition previously discussed in the context of compatibility of quantum measurements \cite{Busch:1986kx, Lahti:2003iw, Son:2005bc, Ali:2009kk, Wolf:2009ik, Yu:2010ji, Heinosaari:2010jj, Banik:2013ei}. To rewrite \cref{eq:NSIT-operators} we use that $\int \dd a\, A_a^\dagger \hat A_a = 1$. This yields
\begin{equation}
  P_{\hat B | \hat A}(b) - P_{\hat B}(b) = \int \dd a\, \tr[ (\hat A_a^\dagger \hat B_b^\dagger \hat B_b \hat A_a - \hat B_b^\dagger \hat A_a^\dagger \hat A_a \hat B_b) \hat \rho_0].
\end{equation}
The trace in the above equation can be interpreted as the expectation value of the Hermitian operator $\int \dd a (\hat A_a^\dagger \hat B_b^\dagger \hat B_b \hat A_a - \hat B_b^\dagger \hat A_a^\dagger \hat A_a \hat B_b)$. For NSIT$_{(0)0}$ to be universally valid, we require that it is zero for \emph{all} initial states $\hat \rho_0$. Thus, the operator itself has to be zero,
\begin{equation}
  \label{eq:nsit-operators-00}
  \begin{split}
    \forall &\hat \rho_0\!: \text{NSIT}_{(0)0} \\
                         &\Leftrightarrow~ \forall b\!: \int \dd a\, (\hat A_a^\dagger \hat B_b^\dagger \hat B_b \hat A_a - \hat B_b^\dagger \hat A_a^\dagger \hat A_a \hat B_b) = 0.
  \end{split}
\end{equation}
This equation can be further simplified to $\int \dd a\, \hat A_a^\dagger \hat B_b^\dagger \hat B_b \hat A_a = \hat B_b^\dagger \hat B_b$. Note that for Hermitian operators $\hat A_a = \hat A_a^\dagger$, $\hat B_b = \hat B_b^\dagger$ we can rewrite \eqref{eq:nsit-operators-00} using the commutator
\begin{equation}
  \forall \hat \rho_0\!: \text{NSIT}_{(0)0} ~\Leftrightarrow~ \forall b\!: \int \dd a\, [\hat A_a \hat B_b, \hat B_b \hat A_a] = 0.
\end{equation}
Furthermore, we have as sufficient conditions the vanishing commutators
\begin{equation}\label{eq:commutator-abba}
  \forall a, b\!: [\hat A_a \hat B_b, \hat B_b \hat A_a] = 0 \Rightarrow~ \forall \hat \rho_0\!: \text{NSIT}_{(0)0},
\end{equation}
and, consequently,
\begin{equation}\label{eq:commutator-ab}
  \forall a, b\!: [\hat A_a, \hat B_b] = 0 \Rightarrow~ \forall \hat \rho_0\!: \text{NSIT}_{(0)0}.
\end{equation}
It is interesting to note that both of these commutator conditions are, generally, only \emph{sufficient} but \emph{not necessary} for NSIT$_{(0)0}$. In fact, a formulation of NSIT$_{(0)0}$ must inherently have an asymmetry \cite{Heinosaari:2010jj} between the test and final measurements, but both \eqref{eq:commutator-abba} and \eqref{eq:commutator-ab} are symmetric under exchange of $\hat A$ and $\hat B$ \footnote{A simple example for this are the Pauli matrices with $\hat A = \hat \sigma_x, \hat B = \hat \sigma_y$. Then, $[\hat A, \hat B] = 2 i \hat \sigma_z$ and $[\hat A \hat B, \hat B \hat A] = 0$. Although the first commutator is non-zero, NSIT$_{(0)0}$ is trivially fulfilled. The physical interpretation of a $\hat \sigma_x$ measurement (or rather, its corresponding POVM element $\mathds 1$) is a single-qubit operation without a meaningful measurement outcome.}.

We can, however, show that vanishing commutators in \eqref{eq:commutator-abba} and \eqref{eq:commutator-ab}, are sufficient and necessary when $\hat A_a, \hat B_b$ are von Neumann projective measurements ($\hat A_a^2 = \hat A_a, \hat B_b^2 = \hat B_b$). Let us start by rewriting the equality in \eqref{eq:nsit-operators-00} using $\hat A_a = \ketbra{a}{a}$ and $\hat B_b = \ketbra{b}{b}$:
\begin{equation}\label{eq:proj-b}
  \int \dd a\, |\langle a|b \rangle|^2 \ketbra{a}{a} = \ketbra{b}{b}.
\end{equation}
Since $\ketbra{b}{b}$ is a projector, squaring the integral on the left-hand side must leave it unchanged. Using the fact that in order to sum up to identity, the $\hat A_a$ must be orthogonal projectors, and therefore $\braket{a|a'} = \delta(a - a')$, we obtain
\begin{equation}\label{eq:proj-b-squared}
  \left[\int \dd a\, |\langle a|b \rangle|^2 \ketbra{a}{a}\right]^2
  = \int \dd a\, |\langle a|b \rangle|^4 \ketbra{a}{a}.
\end{equation}
Comparing \cref{eq:proj-b} and \cref{eq:proj-b-squared}, we see that $|\langle a|b \rangle|^2 = |\langle a|b \rangle|^4$ can only be fulfilled if it is non-zero for exactly one $a$. Thus, $\ket{b}$ is an eigenstate of $\hat A_a$, and the commutator is $[\hat A_a, \hat B_b] = 0$. We have therefore demonstrated that for von Neumann measurements (but not for general POVMs), vanishing commutators in \eqref{eq:commutator-abba} and \eqref{eq:commutator-ab} are both sufficient and necessary for NSIT$_{(0)0}$.

\subsection{NSIT with time evolution}

Let us now consider NSIT$_{(0)T}$ with unitary time evolution $\hat U = e^{-i \hat H t}$. Analogous to the derivation of \eqref{eq:nsit-operators-00} and defining $\tilde B_b^T \equiv \hat U_T^\dagger \hat B_b \hat U_T$, we obtain
\begin{equation}
  \label{eq:nsit-operators-0T}
  \begin{split}
    \forall &\hat \rho_0\!: \text{NSIT}_{(0)T} \\
            &\Leftrightarrow~ \forall b\!: \int \dd a\, (\hat A_a^\dagger (\tilde B^T_b)^\dagger \tilde B^T_b \hat A_a - (\tilde B^T_b)^\dagger \hat A_a^\dagger \hat A_a \tilde B^T_b) = 0,
  \end{split}
\end{equation}
and, if $\hat A_a, \hat B_b$ are Hermitian operators,
\begin{equation}
  \forall \hat \rho_0\!: \text{NSIT}_{(0)T} ~\Leftrightarrow~ \forall b\!: \int \dd a\, [\hat A_a \tilde B_b^T, \tilde B_b^T \hat A_a] = 0.
\end{equation}
Comparing \eqref{eq:nsit-operators-00} and \eqref{eq:nsit-operators-0T}, we can apply the results for NSIT$_{(0)0}$ derived above, namely
\begin{equation}\label{eq:commutator-t-abba}
  \forall a, b\!: [\hat A_a \tilde B_b^T, \tilde B_b^T \hat A_a] = 0 ~\Rightarrow~ \forall \hat \rho_0\!: \text{NSIT}_{(0)T},
\end{equation}
and
\begin{equation}\label{eq:commutator-t-ab}
  \forall a, b\!: [\hat A_a, \tilde B_b^T] = 0 ~\Rightarrow~ \forall \hat \rho_0\!: \text{NSIT}_{(0)T}.
\end{equation}
Furthermore, one obtains
\begin{equation}
  \forall a, b\!: [\hat A_a, \hat B_b] = [\hat A_a, \hat U_T] = 0 ~\Rightarrow~ \forall \hat \rho_0\!: \text{NSIT}_{(0)T}.
\end{equation}
If $\hat A_a, \hat B_b$ are von Neumann operators, we have $(\tilde B^T_b)^2 = \hat U_T^\dagger \hat B_b \hat U_T^{} \hat U_T^\dagger \hat B_b \hat U_T^{} = \hat U_T^\dagger \hat B_b \hat U_T^{} = \tilde B^T_b$. Thus, the results from \cref{sub:nsit-no-time} apply here too: For projectors (but not for general POVMs), vanishing commutators in \eqref{eq:commutator-t-abba} and \eqref{eq:commutator-t-ab} are sufficient and necessary for NSIT$_{(0)T}$.

The above results show that a non-classical ``resource'' is required for an experimental violation of NSIT, namely either highly non-classical states (equivalent to non-classical measurements used in their preparation) or non-classical Hamiltonians (usually requiring an extremely large experimental ``control precision'' as discussed in \cite{Jeong:2014hw, Wang:2013ir, Sekatski:2014ej}).

\section{Classicality}
\label{sec:classicality}

As we have indicated in the introduction, the coarse-graining of ``sharp'' quantum measurement operators into ``fuzzy'' classical measurements, plays a crucial role in the transition from quantum mechanics to classical physics \cite{Kofler:2007bd}. However, not every coarse-grained operator can be called classical. As an example, the parity operator (e.g.\ for large spins or photonic states) only differentiates two macrostates, but is in fact highly non-classical. Generally speaking, a suitable coarse-graining should ``lump'' together neighboring eigenvalues, independent of a (quantum) experiment's Hamiltonian. However, Hilbert spaces in quantum mechanics possess no inherent measure for the distance between orthogonal states. Such a measure must thus arise solely out of interaction Hamiltonians. Effectively, any definition of classicality must therefore depend on Hamiltonians spontaneously realized by nature, which define a natural order and closeness of states. In the following, this closeness is established with an \emph{a priori} choice of suitable reference operators. With this reference set, we can write a definition for \emph{classical operators} and \emph{classical Hamiltonians}:
\begin{enumerate}[(I)]
  \item \label{def:operator} A measurement operator is called \emph{classical} with respect to a reference set iff it fulfills the equality in \eqref{eq:nsit-operators-00} pairwise with every member of the set.
  \item \label{def:hamiltonian} A Hamiltonian is called \emph{classical} with respect to a reference set iff the equality in \eqref{eq:nsit-operators-0T} is fulfilled for each combination of measurement operators from the set.
\end{enumerate}

A natural choice for the reference set are coarse-grained versions of quantum operators in phase space. Phase space inherently involves the necessary definition of closeness in a suitable and intuitive way. Several exemplary candidates for different experiments are discussed in the next section.

\section{Classicality of quantum measurements}
\label{sec:examples}

In the following, we will apply our results to a number of physical systems. We will focus on the classicality of operators---condition \eqref{def:operator} from the previous section---and always assume either an immediate test measurement, or free time evolution in between. To measure the overlap of the undisturbed \eqref{eq:P-unmeasured} and the disturbed \eqref{eq:P-measured} probability distributions, we make use of the Bhattacharyya coefficient \cite{Bhattacharyya:1943ux}, as defined by
\begin{equation}\label{eq:overlap}
  V = \int \dd b \, \sqrt{P_{\hat B}(b) P_{\hat B|\hat A}(b)} \in [0, 1].
\end{equation}
The extreme cases of $V = 0$ and $V = 1$ correspond to orthogonal and identical probability distributions, respectively.

\subsection{Quadrature measurements}

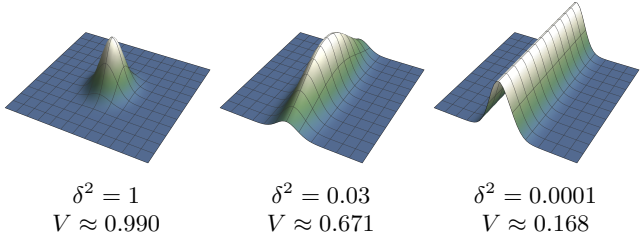
\begin{figure}[t]
  \begin{tikzpicture}
    \node at (0.66\columnwidth, 0) {\includegraphics[width=0.35\columnwidth]{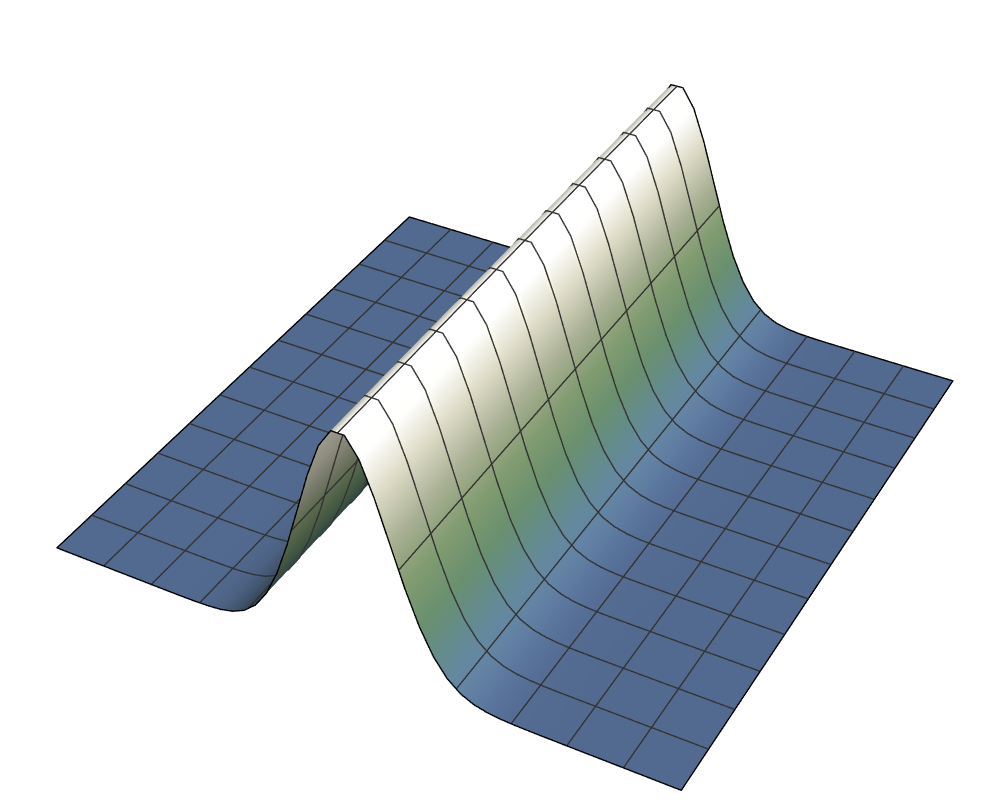}};
    \node at (0.66\columnwidth, -1.8) {$\begin{matrix}\delta^2 = 0.0001 \\ V \approx 0.168\end{matrix}$};
    \node at (0.33\columnwidth, 0) {\includegraphics[width=0.35\columnwidth]{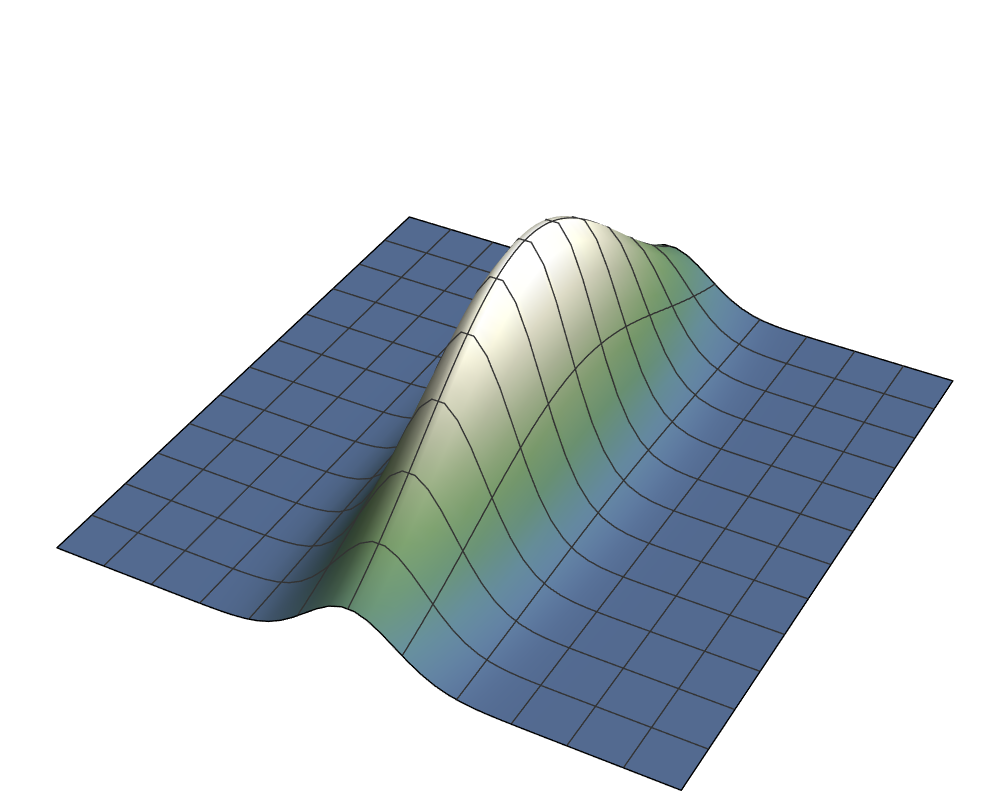}};
    \node at (0.33\columnwidth, -1.8) {$\begin{matrix}\delta^2 = 0.03 \\ V \approx 0.671\end{matrix}$};
    \node at (0.00\columnwidth, 0) {\includegraphics[width=0.35\columnwidth]{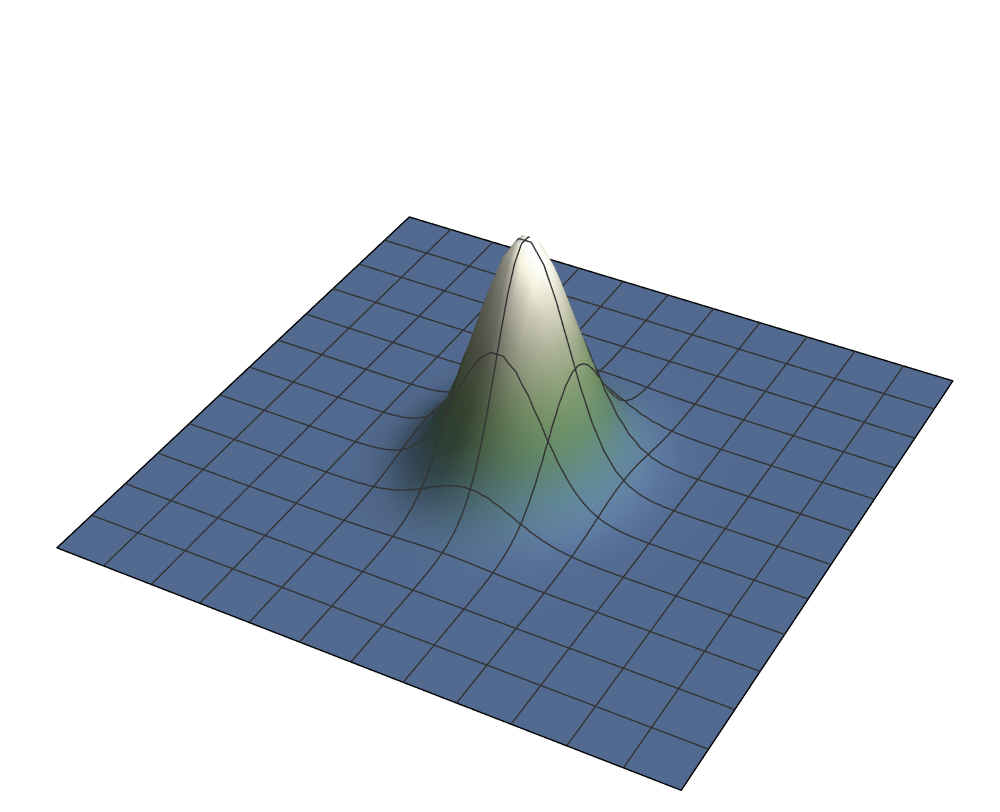}};
    \node at (0.00\columnwidth, -1.8) {$\begin{matrix}\delta^2 = 1 \\ V \approx 0.990\end{matrix}$};
  \end{tikzpicture}
  \caption{\label{fig:p-quadratures}(Color online) Husimi distribution in the complex plane (mesh with interval 1), immediately after a quadrature measurement with decreasing unsharpness $\delta$. Sharp measurements (small $\delta$) completely destroy the initial state, while unsharp measurements (large $\delta$) keep it intact.}
\end{figure}

Let us start with quadrature measurements on pure coherent initial states $\hat \rho = \ketbra{\gamma}{\gamma}$. We investigate coarse-grained measurements with unsharpness $\delta$ in the $X$-quadrature, and unsharpness $\kappa$ in the $P$-quadrature, as described by the (dimensionless) operators
\begin{align}
  \label{eq:cg-x} \hat X^\delta_x &= \frac{1}{(\delta^2 \pi)^{1/4}} \exp\!\left( - \frac{1}{2 \delta^2} (x - \hat X)^2 \right)\!, \\
  \label{eq:cg-p} \hat P^\kappa_p &= \frac{1}{(\kappa^2 \pi)^{1/4}} \exp\!\left( - \frac{1}{2 \kappa^2} (p - \hat P)^2 \right)\!.
\end{align}
Note that for $\hat B_\beta = \pi^{-1} \ketbra{\beta}{\beta}$, we recover the well-known Husimi $Q$-distribution \cite{Husimi:1940vq}, since $P_{\hat B}(\beta) = \pi^{-2} \tr(\ketbra{\beta}{\beta}\hat \rho_0 \ketbra{\beta}{\beta}) = \pi^{-1} \braket{\beta|\hat \rho_0|\beta} = Q(\beta)$. As an example, choosing $\hat A = \hat X^\delta$ and $\hat B_\beta = \pi^{-1}|\beta\rangle\langle\beta|$, the Husimi distribution $P_{\hat B|\hat A}$ is shown in \cref{fig:p-quadratures} for several values of $\delta$.

The behaviors for different combinations of $\hat A, \hat B \in \lbrace \hat X^\delta, \hat P^\kappa \rbrace$ are printed in \cref{table:xp-cg}, and detailed analytic values for the overlaps are listed in \cref{appendix:quadrature-overlaps}.

The importance of selecting a complete set of classical reference operators becomes clear when looking at different combinations of coarse-grained $\hat X^\delta, \hat P^\kappa$. In particular, even a sharp $X$ measurement is revealed by a second (coarse-grained) $X$ measurement only after time evolution. Therefore, $\hat P^\kappa$ has to be a member of the reference set. On the other hand, a sharp measurement in $P$ can never be detected by another measurement in $P$ under free time evolution $\hat H = \hat P^2/(2m)$. Therefore, $\hat X^\delta$ needs to be a member of the set. For $\hat X^\delta$ and $\hat P^\kappa$ to fulfill the consistency condition, we further require sufficiently large $\delta \gg 1$ and $\kappa \gg 1$, such that $[\hat X^\delta, \hat P^\kappa] \approx 0$.

Using the notation $\hat X_\text{c.g.}$ ($\hat P_\text{c.g.}$) for a sufficiently coarse-grained $X$ ($P$) measurement, and $\hat X_\text{sh.}$ ($\hat P_\text{sh.}$) for a sharp, invasive measurement, we can write some candidate reference sets:
\begin{itemize}
  \item $\lbrace \hat X_\text{c.g.} \rbrace$ and $\lbrace \hat X_\text{sh.} \rbrace$ do not constitute reference sets, since they cannot detect the invasiveness of a $\hat X_\text{sh.}$ measurement.
  \item $\lbrace \hat X_\text{sh.}, \hat P_\text{c.g.} \rbrace$ is not a reference set, since the operators do not fulfill \eqref{eq:nsit-operators-00}.
  \item $\lbrace \hat X_\text{c.g.}, \hat P_\text{c.g.} \rbrace$ is a possible reference set.
\end{itemize}
For further discussion about the joint measurability and coexistence of coarse-grained phase space operators we refer the reader to references \cite{Busch:1996fz, Busch:2009di, Busch:2010jl}.

\begin{table}[t]
  \renewcommand*{\arraystretch}{2.0}
  \begin{tabular}{l|c|c}
                                & $\hat A = \hat X^\delta$                             & $\hat A = \hat P^\kappa$                             \\\hline
    $\hat B = \hat X^\delta$ & \pbox{10cm}{$V(0) = 1$\\$V(T\rightarrow\infty) < 1$} & \pbox{10cm}{$V(0) < 1$\\$V(T\rightarrow\infty) = 1$} \\\hline
    $\hat B = \hat P^\kappa$ & $V(t) = \text{const} < 1$                           & $V(t) = 1$
  \end{tabular}
  \caption[]{\label{table:xp-cg}Overlaps \eqref{eq:overlap} between the invaded and the non-invaded probability distributions with different combinations of coarse-grained phase space quadrature measurements. For final measurements in the momentum quadrature, $\hat B = \hat P^\kappa$, the overlap of the system stays constant, since $\hat P^\kappa$ commutes with the free Hamiltonian. For analytical values and detailed discussion see \cref{appendix:quadrature-overlaps}.}
\end{table}

\subsection{Coherent state measurements}

\begin{figure}[tb]
  \begin{tikzpicture}
    \node at (-3.8, 0) {$V$};
    \node at (0, -2.4) {$d$};
    \node at (0,0) {\includegraphics[width=0.80\columnwidth]{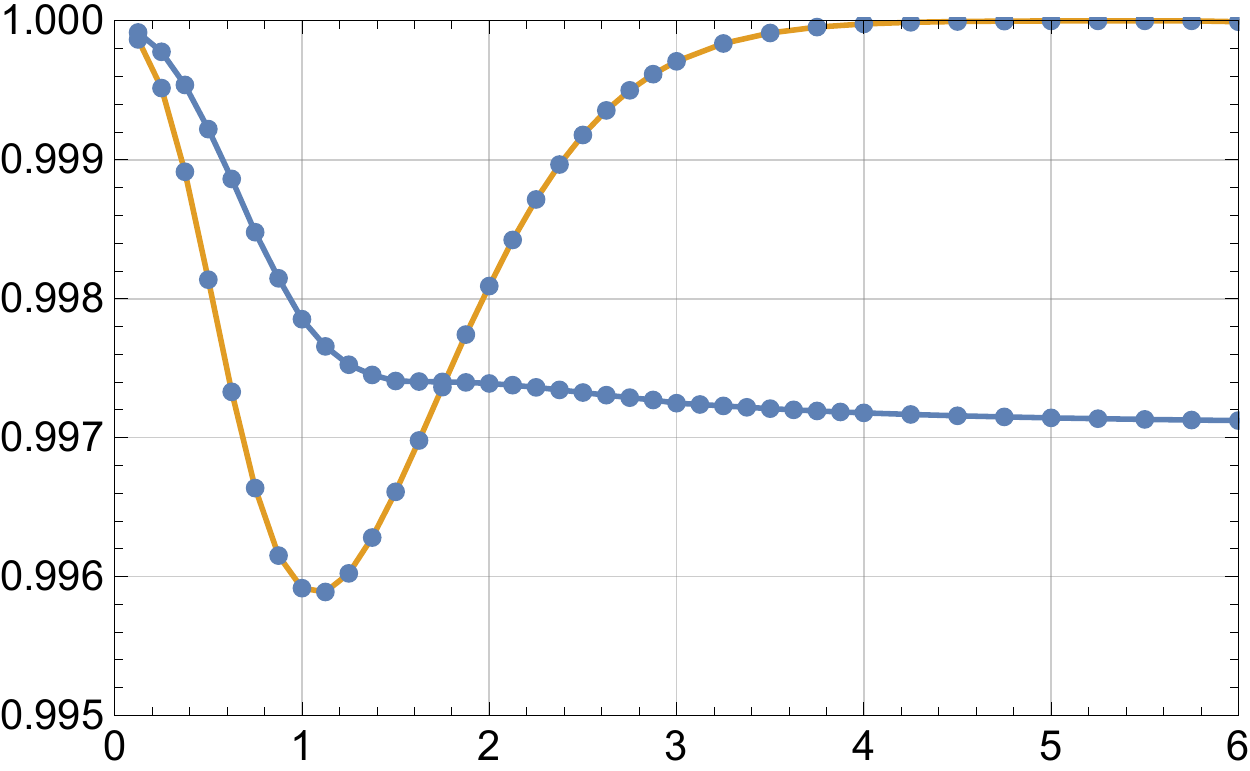}};
  \end{tikzpicture}
  \caption{\label{fig:overlaps-rings}(Color online) Overlap $V$ vs coarse-graining ring width $d$. For coherent initial states in the center of the second region $\ket{\gamma = 3 d/2}$ the overlap approaches unity as more of the state's probability distribution lies in the region. For initial states located on a border $\ket{\gamma = d}$ the overlap approaches a value close to $0.997$. This is due to the artificial sharp boundary between the coarse-grained regions.}
\end{figure}

As another example, let us now consider coarse-grained operators in coherent state space,
\begin{equation}\label{eq:coherent-state-measurement}
  \hat A_a = \frac{1}{\pi} \int \dd \alpha \, f_a(\alpha) \, \ketbra{\alpha}{\alpha},
\end{equation}
where $f_a(\alpha)$ are some real and positive envelope functions that define the coarse-grained regions. Again, we consider coherent initial states $\hat \rho = \ketbra{\gamma}{\gamma}$ and final measurements $\hat B_\beta = \pi^{-1}|\beta\rangle\langle\beta|$. An analytical result can be obtained for a measurement $f_a(\alpha) = \delta(a - \alpha)$ for $a \in \mathbb C$, yielding $\hat A_\alpha = \pi^{-1} \ketbra{\alpha}{\alpha}$. We can now calculate the overlap for $T = 0$:
\begin{equation}
  \begin{split}
    V &= \frac{1}{\pi} \int \dd \beta\, \left[|\langle\beta|\gamma\rangle|^2 \int \dd \alpha\, |\langle\beta|\alpha\rangle\langle\alpha|\gamma\rangle|^2\right]^\frac{1}{2} \\
      &= \frac{2 \sqrt{2}}{3} \approx 0.943.
  \end{split}
\end{equation}
This overlap provides us with a lower bound, that applies to all coarse-grained measurements based on coherent states. As an example, numerically evaluated overlaps for a ring-like coarse-graining ($f_a(r)$ is non-zero for $a d \leq r < (a+1) d$, with $a \in \mathbb N_0$ and $d$ the ring width) are plotted in \cref{fig:overlaps-rings}.

A choice of reference set, alternative to the previously discussed $\lbrace \hat X_\text{c.g.}, \hat P_\text{c.g.} \rbrace$, can be made using the coarse-grained coherent state measurements from \cref{eq:coherent-state-measurement}, i.e.\ $\lbrace \hat A_a \rbrace$ with suitable envelope functions $f_a$ such that $[\hat A_a, \hat A_{a'}] \approx 0$.

\subsection{Fock state measurements}

\begin{figure}[tb]
  \begin{tikzpicture}
    \node at (-3.8, 0) {$V$};
    \node at (0, -2.4) {$\gamma$};
    \node at (0,0) {\includegraphics[width=0.80\columnwidth]{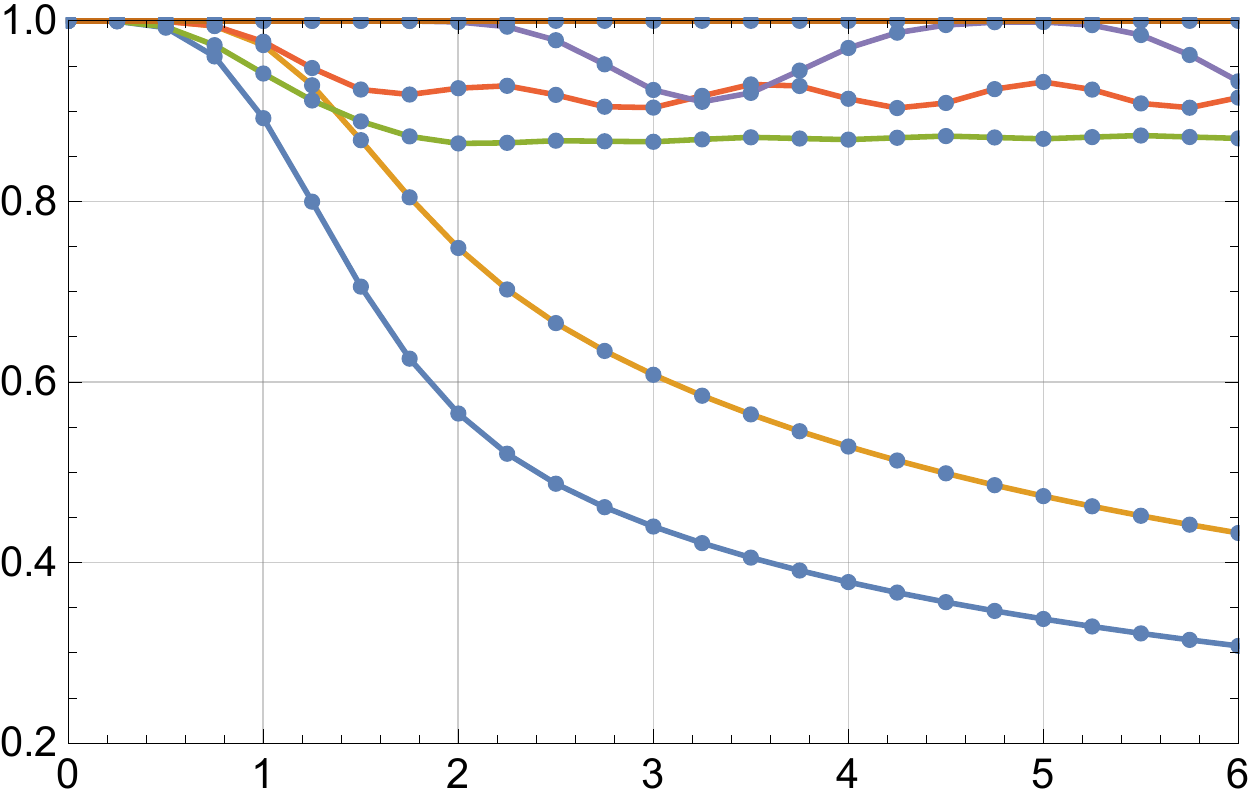}};
  \end{tikzpicture}
  \caption{\label{fig:overlaps-fock}(Color online) Overlap $V$ (cf.\ \cref{eq:overlap}) vs initial state $\ket{\gamma}$ for coarse-grained Fock measurements with different border functions $g(m)$, from top: $100 m^2, 10 m^2, 2 m^2, m^2, 2m, m$. Quadratic border functions are coarse in the coherent state space and therefore not as invasive. Linear border functions lead to increasingly sharp measurements. The oscillations are caused by the fact that the presented type of coarse-graining works better when the initial state is located in the center of a bin. Dips in the overlap occur when the initial state sits at the border between two bins.}
\end{figure}

Instructive examples for observing the effect of coarse-graining are different combinations of Fock-measurements on coherent initial states. We look at coarse-grained von Neumann measurement operators defined by different border functions $g(m)$:
\begin{equation}
  \hat A_m = \sum_k
  \begin{cases}
    \ketbra{k}{k} & \text{if}~ g(m) \leq k < g(m+1), \\
    0 & \text{else}.
  \end{cases}
\end{equation}
For $g(m) = c m^2$ with $c > 0$, the region corresponding to each operator is constant-sized in the coherent state space, since the average photon number is $\bar n = |\alpha|^2$. For sufficiently large $c$ the measurement is therefore sufficiently coarse-grained. Measurements with constant-sized regions in Fock space, $g(m) = c m$, correspond to increasingly sharp measurements in coherent state space. The resulting overlap for different choices of $g(m)$ can be calculated numerically and is discussed in \cref{fig:overlaps-fock}.

\section{Conclusion and outlook}
\label{sec:conclusion}

In contrast to a still widespread belief, we showed that the assumption of macrorealism per se is implied by a strong interpretation of non-invasive measurability. Moreover, no-signaling in time (NSIT), i.e.\ non-invasiveness on the statistical level, is in general a more reliable witness for the violation of macrorealism than the well-known Leggett-Garg inequalities, which are based on two-time correlation functions. In fact, we demonstrated that the right combination of various NSIT conditions serves not only as a necessary but also a sufficient condition for a macrorealistic model for measurements at the predefined time instants accessible in the experiment. We then derived operational criteria for the measurement operators and the system Hamiltonian, whose fulfillment guarantees that no violation of macrorealism can in principle be observed. We argued that these conditions can be used to define the ``classicality'' of measurements, and by extension, of the system's time evolution. Finally, we showed that the classicality of measurements is arbitrarily well fulfilled by suitably coarse-grained versions of quantum measurements.

While our results suggest that an experimental demonstration of non-classicalities requires either very precise measurements or a complex time evolution, a general proof of this trade-off (in terms of experimental control parameters) is still missing. Moreover, coarse-graining, which leads to the classicality of measurements, already requires the notion of ``closeness'' or ``neighborhood'' of eigenvalues, and thereby an understanding of classical phase space. This notion itself stems from Hamiltonians that are spontaneously realized in nature and govern our physical world. The present definition of classicality mitigates this circularity with the choice of an a-priori set of classical measurements. However, it is an open question whether the presupposition of classical phase space can be avoided, or whether it is a fundamental requirement for understanding the quantum-to-classical transition.

\begin{acknowledgments}
  We thank Guido Bacciagaluppi, Gemma de las Cuevas, Owen J.\ E.\ Maroney, and Chris G.\ Timpson for fruitful discussions and helpful comments on the manuscript. We acknowledge support from the European Union Integrated Project Simulators and Interfaces with Quantum Systems.
\end{acknowledgments}

\appendix

\section{Proof that NSIT$_{0(1)2}$ is sufficient for NIC$_{0(1)2}$}
\label{appendix:nsit-nic}

Let us use the short notation $P_i(\pm_i) \equiv P_i(Q_i = \pm)$. Then, the correlations in NIC$_{0(1)2}$ can be written as
\begin{equation}
  \begin{split}
    C_{02} = & + P_{02}(+_0, +_2) + P_{02}(-_0, -_2) \\
             & - P_{02}(+_0, -_2) - P_{02}(-_0, +_2),
  \end{split}
\end{equation}
and, for the variant with a measurement at $t_1$,
\begin{equation}
  \begin{split}
    C_{02|1} = & + P_{012}(+_0, +_2) + P_{012}(-_0, -_2) \\
               & - P_{012}(+_0, -_2) - P_{012}(-_0, +_2).
  \end{split}
\end{equation}
Using NSIT$_{0(1)2}$, i.e.\ $P_{02} (Q_0, Q_2) = P_{012} (Q_0, Q_2)$, we immediately see that NSIT$_{0(1)2}$ is sufficient for $C_{02} = C_{02|1}$, and therefore for NIC$_{0(1)2}$.

\section{Overlaps for quadrature measurements}
\label{appendix:quadrature-overlaps}

In the following we will give analytical values for the overlap for different combinations of coarse-grained $\hat X^\delta$ and $\hat P^\kappa$ measures, as defined by \cref{eq:cg-x} and \cref{eq:cg-p}, acting on a particle with initial state $\braket{x|\psi} = \pi^{-1/4}\sigma^{-1/2}\exp(-x^2/(2 \sigma^2))$. In between the measurements we apply a unitary generated by a free Hamiltonian $\hat U_T = \exp(- i t \hat p^2 / 2m)$. There are four combinations:
\begin{itemize}
  \item $\hat A = \hat X^\delta, \hat B = \hat X^\delta$.
    Here the overlap starts at $V(0) = 1$, but approaches the value
    \begin{equation}
      \lim_{t \rightarrow \infty} V(t) = \frac{4 \delta^2 (\delta^2 + \sigma^2)}{(2 \delta^2 + \sigma^2)^2}.
    \end{equation}
    The effect of the measurement only becomes apparent with time evolution.

  \item $\hat A = \hat P^\kappa, \hat B = \hat X^\delta$.
    The overlap starts at
    \begin{equation}
      V(0) = \frac{4 \kappa^2 (\delta^2+\sigma ^2) [\kappa^2 (\delta^2+\sigma ^2)+1]}{[2 \kappa^2 (\delta^2+\sigma ^2)+1]^2},
    \end{equation}
    and approaches $1$ for $t \rightarrow \infty$. The momentum measurement changes the spatial distribution once, but with wave packet expansion the impact becomes less apparent. 

  \item $\hat A = \hat X^\delta, \hat B = \hat P^\kappa$.
    The overlap is constant in time at the value
    \begin{equation}
      V = \frac{4 \delta^2 (\kappa^2 \sigma ^2+1) [\delta^2 (\kappa^2 \sigma ^2+1)+\sigma ^2]}{[2 \delta^2 (\kappa^2 \sigma ^2+1)+\sigma^2]^2},
    \end{equation}
    since $[\hat P^\kappa, \hat H] = 0$.

  \item $\hat A = \hat P^\kappa, \hat B = \hat P^\kappa$.
    The overlap is constant at $1$, and a measurement in $\hat P$ cannot be detected by a second $\hat P$ measurement, as again, $[\hat P^\kappa, \hat H] = 0$.
\end{itemize}

These examples reaffirm the importance of the selection of multiple final measurements.

\bibliography{macrorealism,arxiv}

\end{document}